\documentclass[%
 reprint,
superscriptaddress,
 amsmath,amssymb,
 aps,
]{revtex4-2}

\usepackage{graphicx}
\usepackage{dcolumn}
\usepackage{bm}
\usepackage{color,soul}
\usepackage{color}
\definecolor{Green}{rgb}{0.10,0.60,0.00}

\begin{document}

\title{Long-range influence of a pump on a critical fluid}

\author{Ydan Ben Dor}
\author{Yariv Kafri}
\affiliation{Department of Physics, Technion -- Israel Institute of Technology, Haifa 3200003, Israel}
\author{David Mukamel}
\affiliation{Department of Physics of Complex Systems, Weizmann Institute of Science, Rehovot 7610001, Israel}
\author{Ari M. Turner}
\affiliation{Department of Physics, Technion -- Israel Institute of Technology, Haifa 3200003, Israel}

\begin{abstract}
A pump coupled to a conserved density generates long-range modulations, resulting from the non-equilibrium nature of the dynamics. We study how these modulations are modified at the critical point where the system exhibits intrinsic long-range correlations. 
To do so, we consider a pump in a diffusive fluid, which is known to generate a density profile in the form of an electric dipole potential and a current in the form of a dipolar field above the critical point. 
We demonstrate that while the current retains its form at the critical point, the density profile changes drastically.
At criticality, in $d<4$ dimensions, the deviation of the density from the average is given by ${\rm sgn}(\cos(\theta))|\cos({\theta})/r^{(d-1)}|^{1/\delta}$ at large distance $r$ from the pump and angle $\theta$ with respect to the pump's orientation. At short distances, there is a crossover to a $\cos({\theta})/r^{d-3+\eta}$ profile. Here $\delta$ and $\eta$ are Ising critical exponents. The effect of the local pump on the domain wall structure below the critical point is also considered. 
\end{abstract}

\maketitle

Consider a fluid continuously pumped by a localized force. When the force couples to a conserved field, such as momentum or density, it leads to a non-local steady-state flow in the system. A canonical example arises when a localized force is exerted on an incompressible fluid at low Reynolds's number. The resulting flow is long-ranged, decaying as a power law with the distance from the pump~\cite{stokes1851effect,batchelor2000introduction}. The solution, known as a Stokeslet, is the Green's function of Stokes's equations. This solution plays an important role in the understanding of many phenomena. Examples include micro-swimmers~\cite{elgeti2015physics,bechinger2016active,kos2018elementary}, hydrodynamic interactions~\cite{brenner1966hydrodynamic,kim2013microhydrodynamics,dufresne2000hydrodynamic,goldfriend2015hydrodynamic}, and the large class of problems associated with slender-body motion~\cite{batchelor1970slender,johnson1980improved,tanasijevic2021hydrodynamic}. In this case the force couples to the momentum flux, since it acts as a source of momentum. 

\begin{figure*}[t]
 	\centering
	\includegraphics[width=\textwidth]{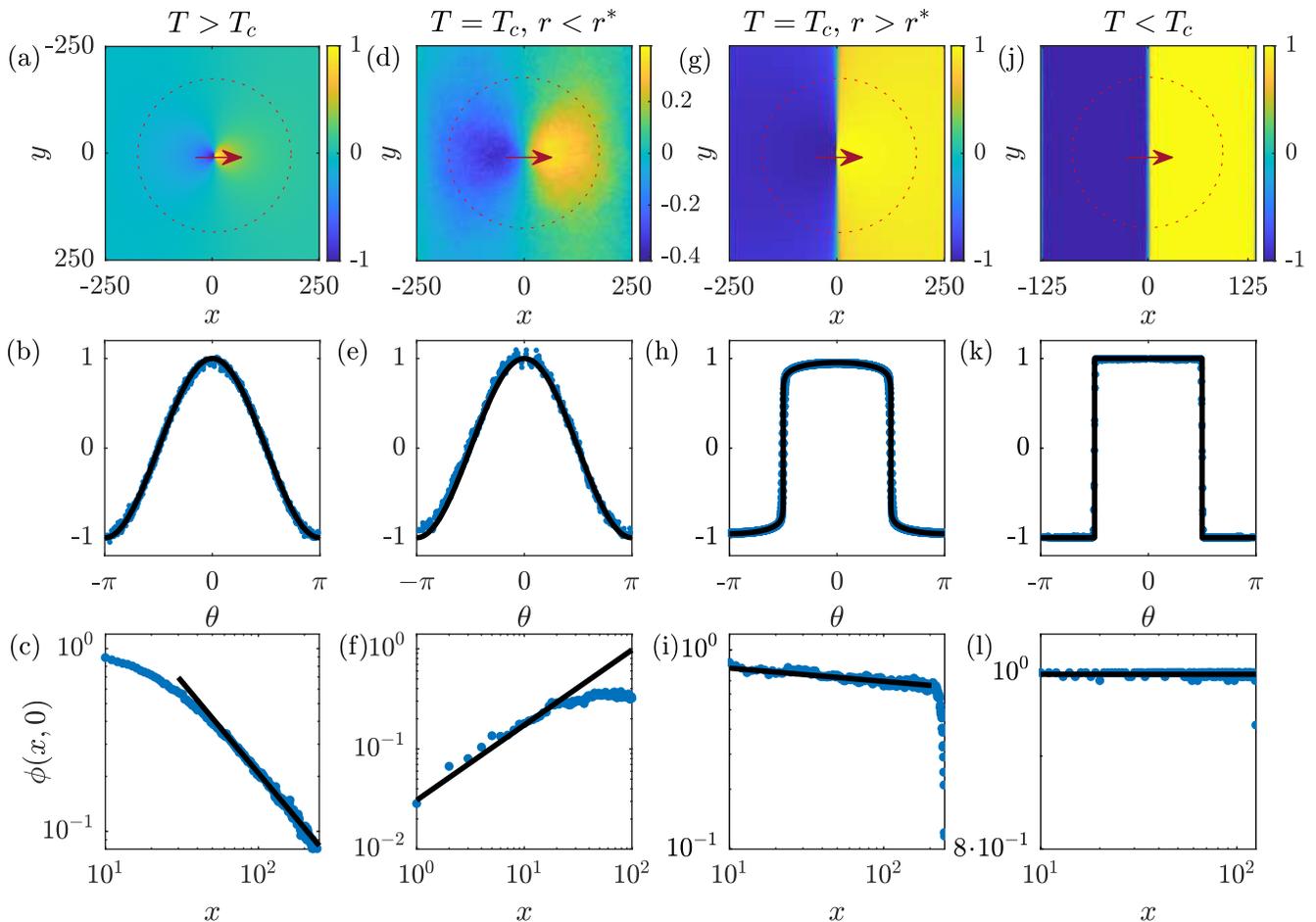}
	\caption{
	Results for a two dimensional $L\times L$ lattice gas with zero magnetization in the presence of a pump (indicated by an arrow) above ($T= 4.54 J$), at ($T = 2.27J$), and below ($T= 1.14 J$) criticality. The top row ((a),(d),(g),(j)) shows magnetization profiles, the middle row ((b),(e),(h),(k)) the measured angular dependence of the density at $r=0.35 L$ (on the dotted circles in the top row) compared to the theoretical prediction (black line, Eqs.~\eqref{eq:above Tc},\eqref{eq:lin fluct},\eqref{eq:nonlin fluct}), and the bottom row ((c),(f),(i),(l)) the  radial dependence of the magnetization along the direction of the pump, as a function the distance from it compared to the theoretical prediction (black line, Eqs.~\eqref{eq:above Tc},\eqref{eq:lin fluct},\eqref{eq:nonlin fluct}). At the critical point we consider two pump strengths allowing us to verify the behavior below and above $r^*$. For $T>T_c$ and $T=T_c\ L=512$ while for $T<T_c\ L=256$.
	}
 	\label{fig:combined}
\end{figure*}

A simpler case is that of a diffusive system where a localized pump drives the particles in a specific direction. Here momentum is not conserved, but the coupling of the pump to the conserved particle density results in long-range currents accompanied by density modulations due to the finite compressibility. In particular, it has been shown that in a hard core gas, corresponding to infinite temperature, the density profile induced by the pump is of the form of a dipole potential and the current is proportional to the gradient of the density~\cite{sadhu2011,sadhu2014long}. 

The above results pertain to systems which without the pump have short range correlations. Here we ask what happens when the underlying fluid is critical, where one expects non-trivial interplay between the long-range correlations of the critical fluid and the long range perturbation induced by the pump. Detailed analysis of this setup is, however, rather involved as it requires going beyond deterministic hydrodynamics due to the large fluctuations and long-range correlations existing in this system.

In this Letter we make a first step towards addressing this problem by studying a pump in a critical diffusive system. This diffusive problem is directly relevant to interacting colloidal particles~\cite{faber2013controlling,maciolek2018collective} which may be studied next to a surface where energy and momentum are not conserved. These can be pumped, for example, by using optical tweezers to bias the motion of colloids along a small segment in a specific direction~\cite{roichman2007colloidal,roichman2008influence,nagar2014collective}.

To study the interplay between the non-local structure induced by the pump and critical correlations we consider a pump in an interacting dissipative system with density conservation, such as a lattice-gas system evolving by Kawasaki dynamics,
which exhibits a liquid-gas phase transition. We show that above the critical temperature as well as at criticality and  slightly below it, the current takes the form of a dipolar electric field.
In contrast, the behavior of the density changes dramatically as a function of temperature. Above the critical temperature the behavior is qualitatively identical to that found in~\cite{sadhu2011,sadhu2014long} for a hard core gas: At large distance $r$ from the pump the density decays to the average density as $\cos(\theta)/r^{(d-1)}$, with $\theta$ the angle measured with respect to the direction of the driving force and $d>1$ the dimension of the system. 
On the other hand, at the critical point the density develops a non-trivial scale-dependent behavior. In particular, in $d < 4$ dimensions the density exhibits a crossover from one scaling form to another at a distance $r^*$ from the pump which varies as an inverse power of the drive.
At distances $r<r^*$, the density profile takes the form $\cos(\theta)/r^{d-3+\eta}$, while for $r>r^*$, it becomes ${\rm sgn}(\cos(\theta))\left|\cos(\theta)/r^{(d-1)}\right|^{1/\delta}$, where $\eta$ and $\delta$ are the Ising exponents with $\eta=1/4$ and $\delta=15$ in $d=2$ and $\eta\simeq3.63\cdot 10^{-2}$ and $\delta\simeq4.79$ in $d=3$.
This implies that in $d=2$ dimensions and at short distances the magnitude of the density modulation {\it grows} with $r$ as $r^{3/4}$, and has a $\cos(\theta)$ angular dependence. On the contrary, in the far field the magnitude of the density modulation decays extremely slowly as a function of $r$, as $r^{-1/15}$. This is accompanied by a change in the angular dependence of the density profile into ${\rm sgn}(\cos(\theta))|\!\cos(\theta)|^{1/15}$. The crossover distance $r^*$  between the two behaviors scales as $f^{-8/7}$, with $f$ the strength of the pump. These results are compared with numerical simulation in Fig.~\ref{fig:combined} where we use a magnetic Ising system corresponding to a lattice gas, so that $s_i=\pm 1$, the magnetization at site $i$, is related to the density through $n_i=(1+s_i)/2$.
The pump locally swaps two pairs of spins~\cite{SM}, such that for each pair $s_L=+1$ on the left and $s_R=-1$ on the right become $s_L=-1$ and $s_R=+1$. A three-dimensional version of Fig.~\ref{fig:combined} can be found in the Supplementary Material~\cite{SM}.
Finally, we also study the system below the critical temperature and show that the pump controls the shape and location of the domain wall between the two phases. This is illustrated for zero average magnetization in Fig.~\ref{fig:combined}. While in the low temperature phase our theoretical arguments hold only in the vicinity of the critical point, they qualitatively agree with the numerics also at lower temperatures.

To obtain these results it is useful to consider a localized pump acting on an Ising lattice gas with a {\it conserved} magnetization field, $\phi({\bf r})$, representing the local deviation of the density from the overall average density. Hereafter we simply refer to $\phi({\bf r})$ as the local density. The Landau-Ginzburg free energy of the gas is given by   
\begin{equation}
     \mathcal{F}_0 = \intop d^dr\,\left[\frac{K}{2}\left|\nabla \phi\right|^2 + \frac{\tau}{2} \phi^2 + \frac{u}{4}\phi^4  \right] \label{eq:F0}\, ,
\end{equation}
with $K,u>0$, and $\tau\propto\left(T-T_c\right)/T_c$. The model evolves by the magnetization-conserving Model B dynamics
\begin{align}\label{eq:EoM}
    \partial_t \phi =& -\nabla \cdot {\bf J} \ , \nonumber \\
     {\bf J} =& {\bf J}_0 + \boldsymbol{\Lambda}  +M{\bf f}\delta^{(d)}\left({\bf r}\right)\ ,
     \end{align}
     where ${\bf J}\left({\bf r}\right)$ is the current. Here ${\bf J}_0$ is the usual deterministic part of the current,
     \begin{equation}
         {\bf J}_0 = -M\nabla\mu\left[\phi\right] = -M\nabla \frac{\delta \mathcal{F}_0}{\delta \phi}\ ,
     \end{equation}
     with $\mu\left[\phi\right]$ the chemical potential.
The Gaussian white noise term, $\boldsymbol{\Lambda}\left({\bf r},t\right)$, has zero mean with a variance satisfying  $\langle\Lambda_i\left({\bf r},t\right)\Lambda_j\left({\bf r}',t'\right)\rangle = 2D\delta_{ij}\delta^{(d)}\left({\bf r}-{\bf r}'\right)\delta\left(t-t'\right)$, where the angular brackets denote an average over histories. 
The pump, of fixed strength ${\bf f}$, localized at the origin, is accounted for by the last term in Eq.~\eqref{eq:EoM}. It is represented by a delta function, which yields the correct behavior in the far field for any localized drive. 
Finally, $D=MT$ with $T$ the temperature, and $M$ is the mobility. 
In general $M$ depends on the magnetization $\phi$. In what follows we consider the case of $\phi$-independent mobility. This is valid above and in the vicinity of the critical point where the coarse grained magnetization is small, so that a small $\phi$ expansion can be applied. Implications of magnetization-dependent mobility are discussed at the end.

We first consider the current. To do so we use density conservation; because of the constant mobility the steady-state average of the chemical potential satisfies Poisson's equation
\begin{equation}\label{eq:Poisson's equation}
	\nabla^2 \langle \mu \rangle = \nabla \cdot \left[{\bf f} \delta^{(d)}\left({\bf r}\right)\right]\ .
\end{equation}
This implies that $\langle \mu \rangle = \frac{1}{S_d}\frac{{\bf f}\cdot{\bf r}}{r^d}$ with $S_d=2\pi^{\frac{d}{2}}/\Gamma\left(\frac{d}{2}\right)$ the area of a $d$-dimensional unit sphere and $\Gamma$ is the Gamma function. Hence the average steady-state current takes the form of the field of an electric dipole:
\begin{equation}\label{eq:above Tc}
	\langle {\bf J} \rangle({\bf r}) = -M\nabla \langle \mu \rangle = \frac{M}{S_d}\frac{1}{r^d}\left[\frac{d\left({\bf f}\cdot{\bf r}\right){\bf r}}{r^2}-{\bf f}\right]\;.
\end{equation}
This motivates us, following existing literature, to refer to $f$ as the dipole strength. This result is confirmed numerically in Fig.~3 of the SM~\cite{SM}. 

Before proceeding to the analysis of the density profile in the various temperature regimes we note that
while the system is \emph{out of equilibrium}, its steady-state properties such as the density profile may be obtained by studying an equivalent {\it equilibrium} system. 
This observation is found useful in the analysis that follows and in the numerical studies of the model. To see this, we use a Helmholtz-Hodge decomposition
\begin{equation}\label{eq:Helmholtz decomposition}
	{\bf f} \delta^{(d)}\left({\bf r}\right) = \nabla h_{\rm eff} + {\boldsymbol \zeta}\ ,
\end{equation}
where $h_{\rm eff}$ is a scalar function and $\boldsymbol{\zeta}$ satisfies
 $\nabla\cdot{\boldsymbol \zeta} = 0$. 
Using this in Eq.~\eqref{eq:EoM} shows that the density profile is only affected by $\nabla h_{\rm eff}$. The statistics of $\phi$ are then described by an {\it equilibrium} problem with the free energy
\begin{align}\label{eq:effectiveh}
	\mathcal{F} =& \mathcal{F}_0 - \intop d^dx\,h_{\rm eff}\phi\ ,
\end{align}
with $\mathcal{F}_0$ given in Eq.~\eqref{eq:F0} and  $h_{\rm eff}$ accounting for the pump. Taking the divergence of Eq.~\eqref{eq:Helmholtz decomposition}, one finds that
\begin{equation}\label{eq:h_eff}
	h_{\rm eff}\left({\bf r}\right) = \frac{1}{S_d} \frac{{\bf f}\cdot{\bf r}}{r^d}\ ,
\end{equation}
is {\it non-local}, decaying as a power law.

The equivalence between the equilibrium and non-equilibrium models for the density is verified numerically in Figs.~4-5 of the SM~\cite{SM}. There we compare conserving, non-equilibrium Kawasaki dynamics for a lattice gas with non-conserving, equilibrium Wolff cluster dynamics~\cite{wolff1988lattice,kent2018cluster} for the density. The Wolff cluster algorithm is much more efficient than the conserving Kawasaki dynamics, as it avoids critical slowing down. This allows us to present results for large systems. In Fig. \ref{fig:combined} we present results of the Wolff algorithm above and at the critical point for two-dimensional periodic lattices of size $512\times512$. Below the critical point, where the mapping to the equilibrium problem is not expected to hold due to the dependence of the mobility on the magnetization, we present results using Kawasaki dynamics for the non-equilibrium model with  smaller systems. We now turn to the analysis of the magnetization profiles.

\noindent \textit{\textbf{A pump above the critical temperature.}} 
Above the critical temperature, namely for $\tau>0$, where the correlation length is finite, one may ignore the  $K |\nabla \phi|^2$ term in \eqref{eq:F0} on length scales larger than the correlation length. In addition, far from the pump, $h_{\rm eff}$ is small so the nonlinear term in \eqref{eq:F0} may be ignored as well. Minimizing the free energy \eqref{eq:effectiveh} yields
\begin{equation}\label{eq:above Tc}
	\langle \phi \rangle \sim \frac{1}{S_d \tau} \frac{{\bf f}\cdot{\bf r}}{r^d}\ .
\end{equation}
This density profile is verified numerically in the leftmost column of Fig.~\ref{fig:combined}. For self-consistency it is straightforward to verify that the terms in $\mathcal{F}_0$ ignored in the derivation make negligibly small contribution to~\eqref{eq:above Tc}  at large distances.
This result has previously been obtained in~\cite{sadhu2011,sadhu2014long} for a lattice gas model of hard core diffusing particles, corresponding to infinite temperature. 

As $\tau$ is lowered to approach the critical point the derivation of \eqref{eq:above Tc} becomes invalid, as manifested by the divergence of the expression of the density profile. This is due to the diverging susceptibility $\chi\propto \tau^{-1}$ (or compressibility of the lattice gas) which is linked to the diverging correlation length.
 
\noindent \textit{\textbf{A pump in a critical system.}}
We start with a mean-field calculation, valid for $d>d_c=4$. To this end, we minimize ${\cal F}$, the effective free energy~\eqref{eq:effectiveh}, to obtain
\begin{equation}\label{eq:EOM T=Tc MF}
	 0 = K\nabla^2 \phi_{\rm MF} - u \phi_{\rm MF}^3 + \frac{1}{S_d} \frac{{\bf f}\cdot{\bf r}}{r^d}\ ,
\end{equation}
where we set $\tau=0$. In the far field we expect the nonlinear contribution to be negligible, leading to
\begin{equation}\label{eq:MF}
  \phi_{\rm MF} \propto {\bf f}\cdot{\bf r}/Kr^{d-2} \;.
\end{equation}
Note that this decays slower than the $\tau>0$ solution Eq.~\eqref{eq:above Tc}, but retains the same angular dependence. The self-consistency of the solution can be checked by comparing the contributions of the nonlinear and linear terms in Eq.~\eqref{eq:EOM T=Tc MF}. One finds that $u\phi_{\rm MF}^3 \ll K \nabla^2 \phi_{\rm MF}$ on distances larger than $r^*\propto\left[uf^2/K^3\right]^{1/[2(d-4)]}$, where $f=|{\bf f}|$. This distance is finite for any dipole strength $\bf f$ in $d>4$ and thus at large distance Eq. \eqref{eq:MF} holds.

Next, we consider the behavior for $d<4$. As the results above show, the density modulations in this pumped critical system are described by an equilibrium model with a magnetic field $h_{\rm eff}({\bf  r})$ which decays as a power law with the distance from the pump. Note that in general the presence of a magnetic field $h$ induces a finite correlation length, $\xi(h)$. To derive $\xi(h)$ we use the equation of state $\phi \sim h^{1/\delta}$ with $h$ small, along with the linear-response relation
\begin{equation}
	\frac{\partial \phi}{\partial h} = \beta \intop d{\bf r }\,G\left({\bf r},h\right)\ ,
\end{equation}
where $\beta$ is the inverse temperature. One finds
\begin{equation}
    \xi(h) \sim h^{-\frac{(\delta-1)}{\delta(2-\eta)}}\;. \label{eq:xih}
\end{equation}
Here $G\left({\bf r},h\right) = \xi^{-(d-2+\eta)}g\left(\frac{r}{\xi}\right)$ is the connected magnetization correlation function and $\eta$ the anomalous exponent associated with the correlation length.

Returning to the problem of a space-dependent magnetic field $h_{\rm eff}$, we reason that if in a certain region the field varies on scales much larger than $\xi(h_{\rm eff})$ the system should behave locally as if it is subject to a constant field $h_{\rm eff}$. Using the equation of state $\phi \sim h_{\rm eff}^{1/\delta}$ with Eq. \eqref{eq:h_eff} for $h_{\rm eff}$ one finds
\begin{equation}\label{eq:nonlin fluct}
	\phi^{\rm nonlin} \propto {\rm sgn}({\bf f}\cdot{\bf r})\left|\frac{{\bf f}\cdot{\bf r}}{r^d}\right| ^{\frac{1}{\delta}}\ .
\end{equation}
The condition for this to be valid is that $r \gg \xi(h_{\rm eff})$ which using Eqs. \eqref{eq:xih} and \eqref{eq:h_eff} gives
\begin{equation}\label{eq:r_* fluct}
	r> r^* \sim	f^{\frac{\delta-1}{d(\delta - 1) - \delta (3-\eta) + 1}}\ \;.
\end{equation}
Using the scaling relation $2-\eta=d(\delta-1)/(\delta+1)$ then gives $r^* \sim f^{-(\delta+1)/\left((\delta+1)-d\right)}$.
In $d=4-\epsilon$ dimensions $\delta=3+\epsilon$ so that
the leading order behavior of $r^*$ is 
\begin{equation}\label{eq:r_* fluct_epsilon}
	 r^* \sim	f^{-2/\epsilon}\ .
\end{equation}
which {\it diverges} as $f \to 0$. Using $\delta=15$ and $\delta\simeq 4.79$ in $d=2$ and $d=3$ respectively~\cite{kardar2007statistical,pelissetto2002critical} one finds  $r^* \sim f^{-8/7}$ and  $r^* \sim f^{-2.08}$, respectively. It is interesting to note that this behavior is different than the one in $d>4$ where the $r^*$ decreases with $f$.

All in all we find that the far-field behavior below $d=4$ is very different from that found for $d>4$. In particular, the density profile is found to decay with the distance in the far field as $r^{-0.4}$ and $r^{-1/15}$ in $d=3$ and $d=2$ respectively. Furthermore, the angular dependence is also substantially modified. It is given by ${\rm sgn}(\cos(\theta))|\!\cos(\theta)|^{0.2}$ and ${\rm sgn}(\cos(\theta))|\!\cos (\theta)|^{1/15}$ for $d=3$ and $d=2$ respectively. 

The fact that $r^*$ increases with decreasing $f$ implies that it is also of interest to study the behavior at distances $r<r^*$. Here the field $h_{\rm eff}$ changes much faster than $\xi(h_{\rm eff})$ so that the previous treatment is not self-consistent. Then one can use perturbation theory to understand how the field influences the system. This gives
\begin{equation}
   \phi^{\rm lin} \equiv \langle \phi ({\bf r})\rangle = \int d{\bf r}'\,G_0({\bf r}-{\bf r'}) h_{\rm eff}({\bf r'}) \;, \label{eq:green}
\end{equation}
where $G_0({\bf r}) \propto 1/r^{d-2+\eta}$ is the renormalized Green's function, so that
\begin{equation}\label{eq:lin fluct}
	\phi^{\rm lin} \propto \frac{{\bf f}\cdot{\bf r}}{r^{d-2+\eta}}\ .
\end{equation}
Note that the dipolar form of $h_{\rm eff}$ ensures that the integral Eq.~\eqref{eq:green} converges.
This is the behavior one might expect naively from Eq.~\eqref{eq:MF} when extended to include the fluctuations below $d_c$. The self-consistency of the linear-response approach can be checked by demanding that the fluctuations in the magnetization are larger than the mean induced magnetization. This justifies the use of the critical Green's function. Since the fluctuations of the magnetization on a scale $r$ are given by $r^{-(d-2+\eta)/2}$ and the mean magnetization is given by Eq. \eqref{eq:lin fluct} one finds that the approach breaks down on a scale $r^* \sim f^{2/(d-4+\eta)}$. This coincides with the scale at which the far-field behavior starts to work, see Eq. \eqref{eq:r_* fluct}. Also, note that the two expressions  $\phi^{\rm lin}({\bf r})$ and $\phi^{\rm nonlin}({\bf r})$ match at $r^*$.
 
Interestingly, for dimensions such that $d+\eta<3$, i.e. two dimensions, the magnitude of the near-field density profile for $r < r^*$ is an {\it increasing} function of $r$. The results at the critical point are verified numerically in Fig.~\ref{fig:combined}(d)-(i) for a lattice gas in $d=2$. 

In Sec.~V of the SM~\cite{SM} we also consider the richer behavior that arises close to but not exactly at criticality, due to the competition between the off-critical correlation length and the one induced by the magnetic field.

\noindent \textit{\textbf{A pump below the critical temperature.}} Here we take $\tau <0$ where in the absence of a pump the system phase separates. Consider first the case studied above where the net magnetization in the system is zero. The magnetization is that of an equilibrium system with an effective magnetic field given by Eq.~\eqref{eq:h_eff} which changes sign on the plane ${\bf f} \cdot {\bf r}=0$.
Since $\tau<0$, even a small field induces a large magnetization so that this plane divides the system between positive and negative magnetization regions.
The two domains are separated by a narrow domain wall whose width is controlled by the correlation length of the system. 
The location and orientation of the domain wall is dictated by the pump.
In addition, the magnetization is essentially constant in the bulk of the phases because the effective magnetic field is small. In Section~IV of the SM~\cite{SM} we generalize the treatment to finite periodic systems with the dipole parallel to one of the periodic axes and to arbitrary values of the magnetization. In the latter case curved or closed domain walls appear.

Far below the critical point the local average magnetization becomes large and the magnetization dependence of the mobility is expected to modify our results. In particular, the mapping to an equilibrium model fails. Nonetheless, we show in Figs.~3 and 5 of the SM~\cite{SM} that our theory gives a qualitatively correct picture both for the magnetization and the current when compared to numerical simulations.

In summary, we studied the behavior of a pump in a conserving lattice gas, or its equivalent magnetic system at various temperatures. The most interesting behavior is found at the critical point, where there is an interplay between the long-range effects generated by the drive and the long-range critical correlations in the system.
The problem studied opens the door to a host of questions. For example, it would be interesting to generalize this problem to fluids belonging to other dynamic and static universality classes, in particular to momentum-conserving ones. These might be realized experimentally using setups similar to that of Ref.~\cite{buttinoni2012active}. In addition, it would be interesting to explore these questions in active systems where it is known that asymmetric objects placed in the fluid act as pumps even in absence of external forces~\cite{galajda2007wall,guidobaldi2014geometrical,baek2018generic,granek2020bodies}.
The closest system to the one considered above is dry scalar matter which has a critical point associated with a motility induced phase separation~\cite{cates2015motility,cates_when_2013,Tailleur2008PRL,buttinoni_dynamical_2013,Redner2013PRL,paliwal_chemical_2018}.
Our method might be used as a possible probe of the universality class of the system, a topic which has been under some debate~\cite{caballero2018bulk,partridge2019critical,maggi2021universality,dittrich2021critical,agranov2021exact}.

We thank Tal Agranov, Anna Frishman, Omer Granek, Asaf Miron, and Sunghan Ro for helpful discussions. The work was supported by a research grant from the Center for Scientific Excellence at the Weizmann Institute of Science. YBD and YK are supported by an Israel Science Foundation grant (2038/21) and an NSF-BSF grant (2016624). AMT is supported by an Israel Science Foundation grant (1939/18).

\bibliography{main}

\end{document}


\title{Supplementary material for ``Long-range influence of a pump on a critical fluid''}

\maketitle

\section{Numerical methods}\label{sec:numerical methods}
In this section we describe the numerical methods of this work.

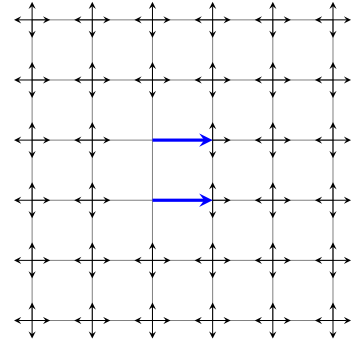
\begin{figure}
    \centering
    \begin{tikzpicture}
        \foreach \i in {\xMin,...,\xMax} {
            \draw [very thin,gray] (\i*0.8,0.8) -- (\i*0.8,4.8);
        }
        \foreach \i in {\yMin,...,\yMax} {
            \draw [very thin,gray] (0.8,\i*0.8) -- (4.8,\i*0.8);
        }
        \foreach \i in {1,...,6} {
            \foreach \j in {1,2,5,6} {
                \draw [stealth-stealth,very thin,black] (\i*0.8-.3*0.8,\j*0.8) -- (\i*0.8+.3*0.8,\j*0.8);
            }
        }
        \foreach \i in {1,2,5,6} {
            \foreach \j in {3,4} {
                \draw [stealth-stealth,very thin,black] (\i*0.8-.3*0.8,\j*0.8) -- (\i*0.8+.3*0.8,\j*0.8);
            }
        }
        \foreach \j in {1,2,5,6} {
            \foreach \i in {1,...,6} {
                \draw [stealth-stealth,very thin,black] (\i*0.8,\j*0.8-.3*0.8) -- (\i*0.8,\j*0.8+.3*0.8);
            }
        }
        \foreach \j in {3,4} {
            \foreach \i in {1,2,4,5,6} {
                \draw [stealth-stealth,very thin,black] (\i*0.8,\j*0.8-.3*0.8) -- (\i*0.8,\j*0.8+.3*0.8);
            }
        }
        \foreach \i in {4} {
            \foreach \j in {3,4} {
                \draw [-stealth,very thin,black] (\i*0.8,\j*0.8) -- (\i*0.8+.3*0.8,\j*0.8);
            }
        }
    
        \draw [-stealth,very thick, blue] (2.4,2.4) -- (3.2,2.4);
        \draw [-stealth,very thick, blue] (2.4,3.2) -- (3.2,3.2);
    \end{tikzpicture}
    \caption{An illustration of the possible exchanges on the two-dimensional lattice, here represented as a lattice gas with $s_i=+1$ for an occupied site and $s_i=-1$ for an empty site. All arrows denote possible exchanges of an occupied site moving into an empty site. Blue arrows stand for transitions associated with the pump. In our numerics these occur whenever an exchange is possible.}
    \label{fig:Lattice}
\end{figure}

\noindent {\bf Kawasaki Dynamics:} To simulate directly the conserved dynamics described in the main text, we use standard local Kawasaki dynamics on a two-dimensional Ising lattice gas of dimensions $L \times L$. The system is initiated with a prescribed total magnetization by setting the relative number of $s=+1$ and $s=-1$ spins. At each Monte Carlo step, a pair of nearest-neighbor spins is chosen at random. If the spins are opposite, they are exchanged with probability $p_{\rm ex} = \min\left[\exp\left(-\beta \Delta E\right),1\right]$ where $\beta$ is the inverse temperature and $\Delta E$ is the energy difference between the exchanged and initial states which is obtained from the Ising energy $E(\lbrace s_i \rbrace) = -J\sum_{\langle i,j \rangle}s_i s_j$.  Here $J>0$ is the coupling constant and $\langle i,j \rangle$ denotes nearest neighbor sites. To enhance the current and density modulation the pumping acts along two bonds at the center of the lattice -- a bond between sites $(L/2,L/2)$ and $(L/2+1,L/2)$ and another bond between sites $(L/2,L/2+1)$ and $(L/2+1,L/2+1)$. The updating rule of the two pump bonds is different. If $s_i=+1$ on the left site and $s_i=-1$ on the right site of either pair of sites they are exchanged. This corresponds to an infinite biasing rate. The possible transitions are depicted in Fig.~\ref{fig:Lattice}.

The Kawasaki dynamics was used to study the density profile below the critical point and the currents in the system across the temperature range, described in the SM below. The current along each bond is measured by counting the exchange events per unit time and the streamlines appearing in Fig.~\ref{fig:current} of the SM were obtained using a standard Matlab tool. In addition, the Kawasaki algorithm was used to test the agreement between the steady-state density with that predicted by the effective equilibrium model.

In panels (j)-(l) of Figure 1 of the main text, the density profile obtained by numerical simulation below the critical point is displayed. It was carried out on an Ising square lattice of size $L\times L$ with $L=256$, imposing helical boundary conditions~\cite{newman1999monte}. This means that the last spin in each row is coupled to the first spin of the next row instead of the first spin of the same row. This is computationally convenient as it allows one to store the lattice in a linear array. To do so we number the lattice sites from $1$ to $L^2$ along the rows, with the first row labelled $1$ to $L$. Then the adjacency rule has a simple description in which the neighbors of the $i^{\rm th}$ spin on the two-dimensional lattice are situated at $(i\pm 1)$ modulo $L^2$ and at $(i\pm L)$ modulo $L^2$.
The density profile was averaged over 300 realizations, running for $1.5\cdot10^6$ Monte Carlo sweeps.

\noindent {\bf Wolff's algorithm:}
Away from the critical point, the system is easily simulated using Kawasaki dynamics. However, most of our work is dedicated to the critical behavior of the system, where it is subject to critical slowing down~\cite{kardar2007statistical}. Indeed, attempts to simulate large systems using Kawasaki dynamics proved to converge extremely slowly, as expected.

To obtain the density profiles in reasonable time, we use the fact that as long as $T$ is not far below the critical temperature, the density profile can be obtained from a mapping onto an effective equilibrium system, as discussed in the main text. The pump used in the Kawasaki dynamics is then replaced by a long-ranged effective magnetic field of strength $f$. This furthermore allows us, through equivalence of ensembles, to use non-conserving dynamics to reach the same equilibrium density profile as that obtained using conserving dynamics.
Therefore, equilibrium, non-conserving, cluster algorithms can be used to measure the desired density profiles at criticality.
In our case, we used an adaptation of Wolff's algorithm~\cite{wolff1988lattice} that allows for a (spatially-varying) magnetic field.  This is done by using a ``ghost spin'' -- an additional degree of freedom that is assigned to the magnetic field itself, as detailed in Ref.~\cite{kent2018cluster} and outlined below. 

The energy function used in the simulations is given by
\begin{equation}
    E(\lbrace s_i \rbrace,s_g) = -J\sum_{\langle i,j \rangle} s_i s_j - \sum_i H_i s_g s_i\;,
\end{equation}
with $s_i,s_j\in\{-1,1\}$ and $H_i$ is the value of the field on the $i^{\rm th}$ site. The difference with the ordinary Ising Hamiltonian is the addition of the ghost spin $s_g\in\{-1,1\}$ which couples to all the spins in the system and sets the global direction of the magnetic field. With it, the $\mathbb{Z}_2$ symmetry of the system in the absence of a magnetic field is restored, at the expense of measuring the state of the system relative to the fluctuating direction set by the ghost spin.
For completeness, we briefly outline the algorithm~\cite{kent2018cluster} used in our simulation. In addition to the spin configuration the algorithm keeps track of two lists of spins. The first is called a cluster and the second a queue. The algorithm repeatedly builds clusters, flips them, and restarts them. The queue is used to keep track of spins that are in the process of being added to the cluster. Each cycle is carried out as follows:
\begin{enumerate}
    \item Start with an empty cluster and an empty queue. Then choose a spin (ordinary or ghost) randomly and add it to the queue.
    \item Until the queue becomes empty:
    \begin{enumerate}
        \item Pick the oldest spin still in the queue, say $s_i$.
        \item If the $i^{\rm th}$ spin $s_i$ is not registered in the cluster,
        \begin{enumerate}
            \item Register it to the cluster.
            \item If the $i^{\rm th}$ spin is an ordinary spin,
            \begin{enumerate}
                \item Add the ghost spin $s_g$ to the queue with probability $p_H = \max\left[1-\exp\left(-2\beta H_i s_g s_i\right),0\right]$. 
                
                \item Add each of its nearest neighbors to the queue with probability $p_s = 1-\exp\left(-2\beta J\right)$ only if the neighboring spin has the same spin value as the $i^{\rm th}$ spin.  
                
            \end{enumerate}
            
            \item If the $i^{\rm th}$ spin is the ghost spin $s_g$ add each spin $j$ of the lattice with probability $p_H = \max\left[1-\exp\left(-2\beta H_j s_g s_j\right),0\right]$ to the queue.
            
        \end{enumerate}
        \item  Remove the spin from the queue.
    \end{enumerate}
    \item After the queue becomes empty flip all the spins in the cluster and return to 1.
\end{enumerate}
Finally, measuring the density involves multiplying the values of all spins by that of the ghost spin, $s_g$.

In the figure of the main text, the simulations above and at criticality were done using the Wolff algorithm presented here. The Ising system was simulated on a $512\times512$ square lattice, imposing helical boundary conditions, as in the Kawasaki simulations.
The temperature of the simulation above the critical point was chosen to be $T=2T_c$, with $T_c\simeq 2.27 J$.
The force strength $f$ was set to $f=1$ for $T>T_c$, $f=0.03$ for $T=T_c$ probing the near-field behavior, and $f=100$ for $T=T_c$ probing the far-field behavior.
The density profile above $T_c$ and the far-field profile at $T_c$ were averaged over 250 realizations, each simulated for $10^3$ cycles. For the near-field density profile at criticality, the average was done over 500 realizations, each simulated for $1.5\cdot 10^4$ cycles. Note that due to the ghost spin, each cycle involves all spins in the lattice and hence analogous to a usual Monte Carlo sweep.

\begin{figure}
 	\centering
	\includegraphics[width=8.6 cm]{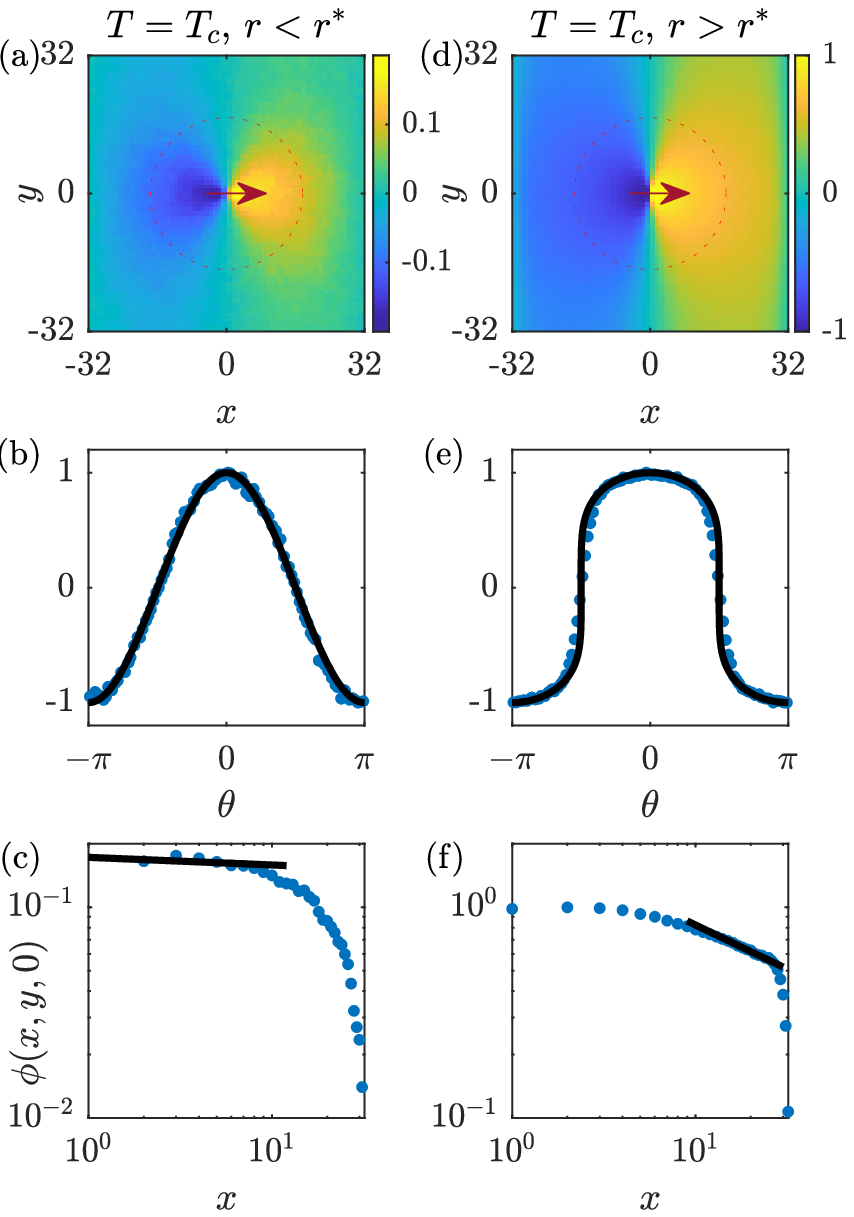}
	\caption {Results for a three dimensional $L\times L\times L$ lattice gas of zero magnetization with $L=64$, in the presence of a pump (indicated by a red arrow) at criticality ($T \simeq 4.51J$). The top row ((a),(d)) shows a slice of the magnetization profiles in the plane of the pump $z=0$, the middle row ((b),(e)) the measured angular dependence of the density slice at $r=0.28 L$ (marked by the dotted circles in the top row) compared to the theoretical prediction (black line), and the bottom row ((c),(f)) the radial dependence of the magnetization along the direction of the pump, as a function the distance from it compared to the theoretical prediction (black line). We consider two pump strengths allowing us to verify the behavior below and above $r^*$ (see text). 
	}
 	\label{fig:3d}
\end{figure}

Figure~\ref{fig:3d} shows the critical behavior of a three-dimensional Ising system. These simulations were done using the same Wolff algorithm on a $64\times64\times64$ cubic lattice. For simplicity, here we used the magnetic field of an infinite system (as given by Eq.~(8) in the main text) and do not account for the periodic boundary conditions. The temperature was set to the numerically-estimated critical value for the three-dimensional cubic Ising model~\cite{ferrenberg2018pushing}, with $\beta_c \simeq 0.22165 J^{-1}$ the inverse critical temperature and $J$ the Ising coupling constant. The pump force strength $f$ was set to $f=0.5$ for probing the near-field behavior and to $f=50$ for probing the far-field behavior. The density profiles were averaged over 100 realizations, simulated for $5\cdot10^3$ cycles (analogous to a Monte Carlo sweep) for the near-field density profile and for $10^3$ cycles for the far-field profile. The results are compared with the theoretical predictions for $r<r^*$ where $\phi\sim \vec{f}\cdot\vec{r}/r^{(d-2+\eta)},\,\eta\simeq0.036$ and for $r>r^*$ where $\phi\sim |\vec{f}\cdot\vec{r}/r^{d}|^{1/\delta},\,\delta\simeq4.79$. Note that here, in contrast to two dimensions, the theory predicts a decaying density profile for $r<r^*$. This is consistent with the numerical results.

\section{Current measurement}
As argued in the main text a localized drive generates a current whose form is that of a field of an electric dipole, and given by Eq.~(5) above, at criticality, and slightly below it.

\begin{figure}
 	\centering
	\includegraphics[width=8.6 cm]{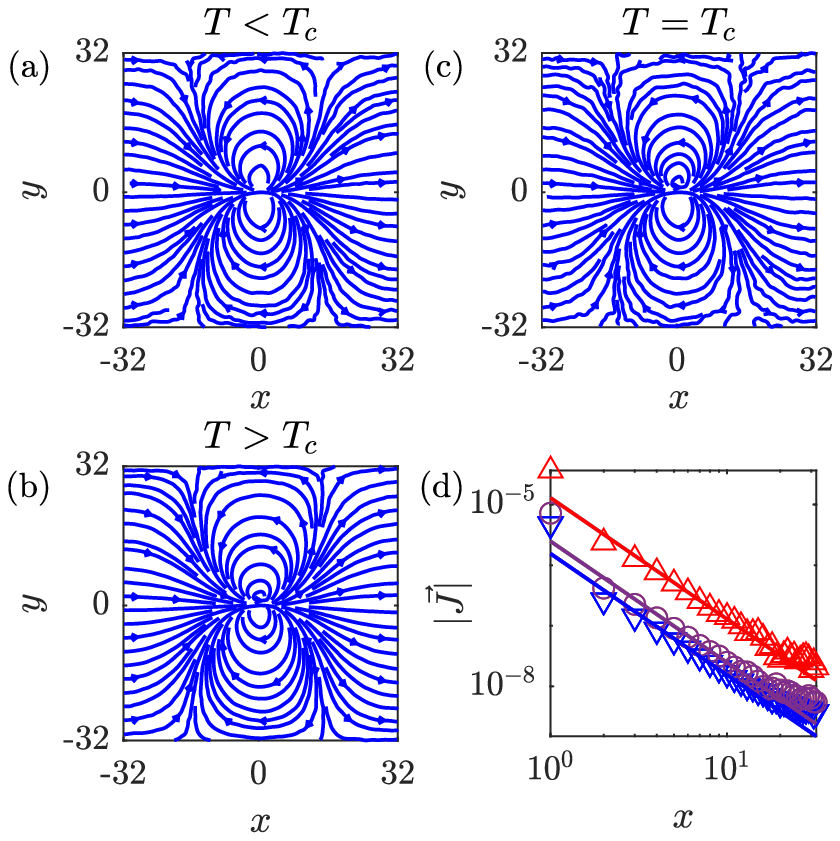}
	\caption{Streamlines and decay of the current induced by a localized pump in $d=2$ (a) below the critical point with $T = 0.9T_c$, (b) at criticality with $T = T_c$, and (c) above the critical point with $T = 2T_c$.  (d) Comparison of the decay of the current amplitude in the direction of the pump below the critical point (blue), at criticality (purple) and above the critical point (red) with the expected $r^{-2}$ power law. 
	}
 	\label{fig:current}
\end{figure}

To verify this, we use the Kawasaki dynamics described above. 
The resulting current streamlines, as well as the $r^{-2}$ decay, at different temperatures above, at, and below the critical point are displayed in Fig.~\ref{fig:current}.
Note that since the pump's strength $f$ depends implicitly on the density of particles at the origin, the amplitude of the current varies as a function of the temperature as well.
Surprisingly, while the mapping to effective equilibrium is not expected to hold far below the critical point, due to the density dependence of the mobility, we find that the current nonetheless exhibits the same dipolar form as if the mobility were constant.

\section{Validity of the effective equilibrium approach}

\begin{figure}
 	\centering
	\includegraphics[width=8.6 cm]{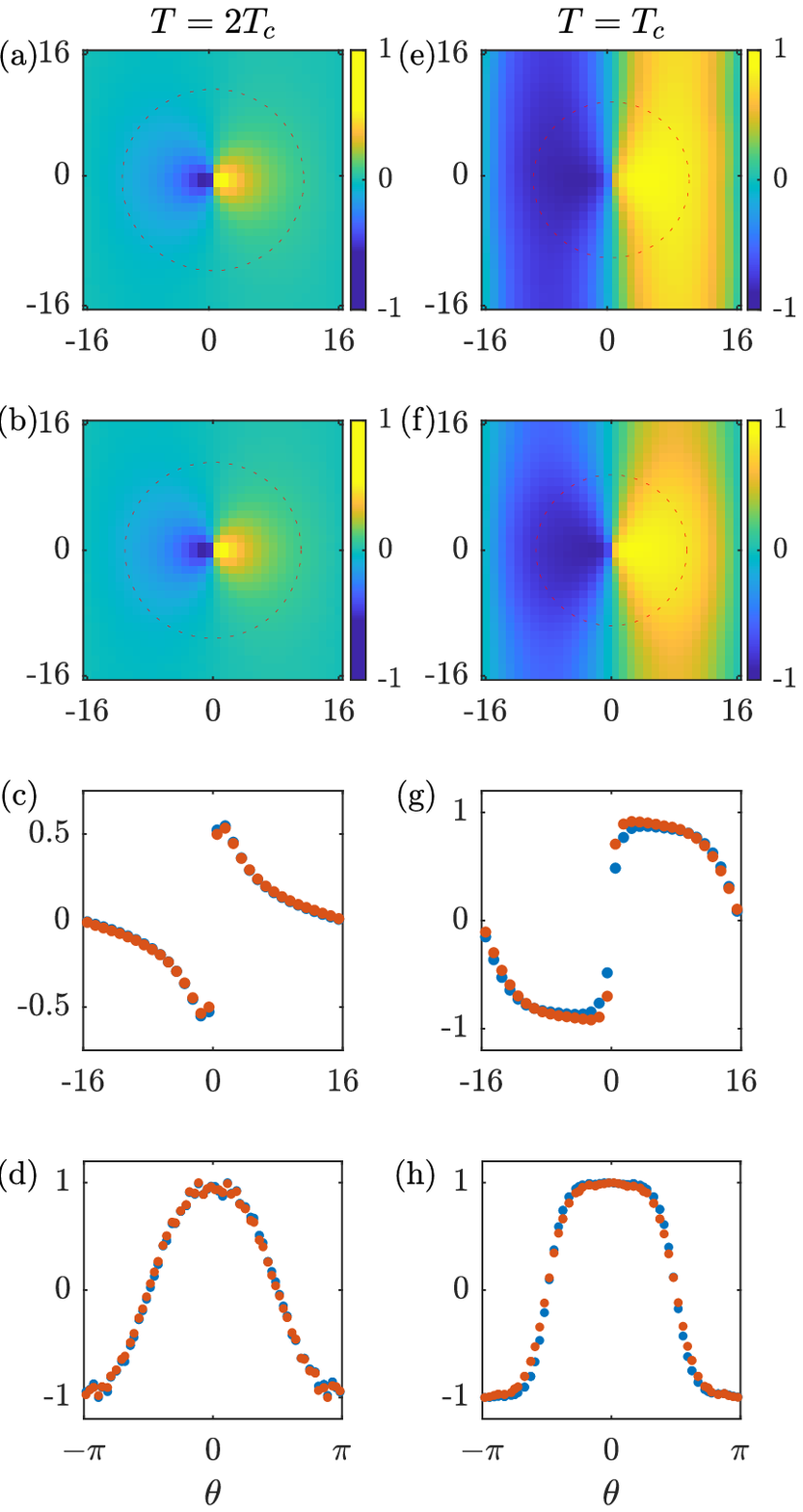}
	\caption{Comparison of density profiles above and at the critical point with different algorithms.
	The left column (a)-(d) shows the density above the critical point $T=2T_c$, whereas the right column (e)-(h) shows its behavior at criticality $T=T_c$.
	The top row ((a),(e)) shows the density profiles measured using the Kawasaki protocol with zero average magnetization. The second row, ((b),(f)) shows the density profiles measured using the Wolff protocol with no offset magnetic field $h_0=0$. For the system above the critical point (b), the effective magnetic field strength was set to $f=0.2$, while in the critical point (e) it was set to $f=1.25\cdot10^{-1}$.
	The third row ((c),(g)) compares slices of the density profiles along $\hat{x}$ for $y=0$, using the Kawasaki (blue) and Wolff (orange) protocols. (c) compares the density profiles shown in (a) and (b), while (g) compares (e) with (f).
	The bottom row, ((d),(h)) compares the angular dependence of the densities measured on the red dashed circles appearing in the first two rows, using the Kawasaki (blue) and Wolff (orange) protocols. (d) compares (a) with (b), measured on a circle of radius $R=0.35 L$, while (h) compares (e) with (f), measured on a circle of radius $R=0.3 L$. In both cases $L=32$.}	
 	\label{fig:comparison1}
\end{figure}

\begin{figure}
 	\centering
	\includegraphics[width=8.6 cm]{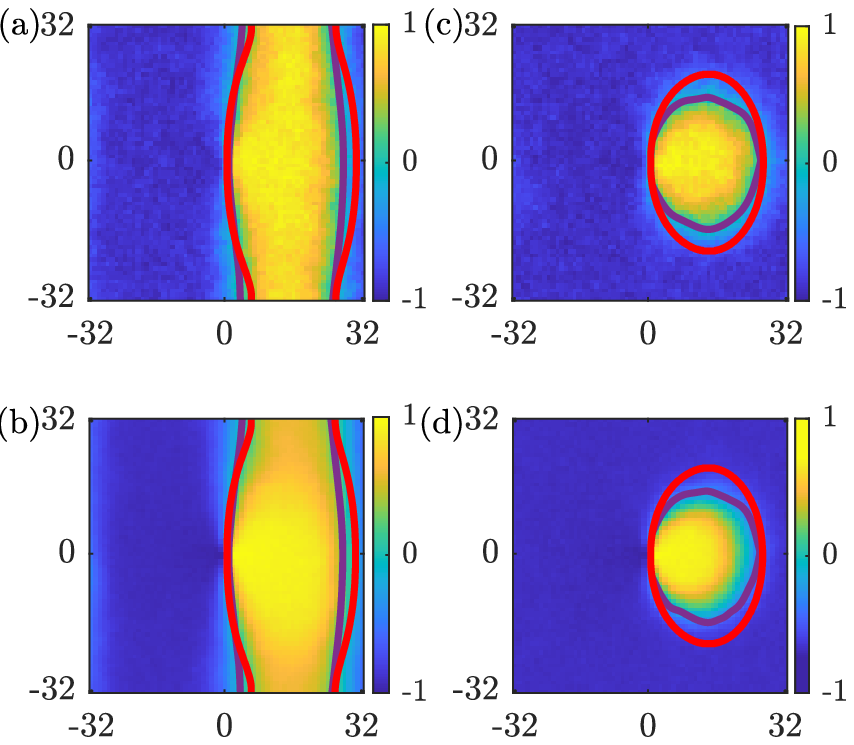}
	\caption{Comparison of the domain walls created in the phase-separated system, below the critical point $T=0.9T_c$. The domain walls predicted by the theory for a periodic lattice, Eq.~\eqref{eq:phase boundary} and~\eqref{eq:magnetic field periodic}, are displayed in red. The measured domain walls appear in purple.
	(a) Density profile using the Kawasaki protocol for an overall magnetization $\langle\phi\rangle = 0.4$.
	(b) Density profile using the Wolff protocol using the effective magnetic field of strength $f=0.08$ and an offset magnetic field $h_0=1.6\cdot 10^{-3} f/a$, with $a$ the lattice constant.
	(c) Density profile using the Kawasaki protocol for an overall magnetization $\langle\phi\rangle = 0.2$.
	(b) Density profile using the Wolff protocol using the effective magnetic field of strength $f=0.08$ and an offset magnetic field $h_0=4.7\cdot 10^{-3}f/a$, with $a$ the lattice constant. Note that despite the magnitude of the average \emph{local} magnetization in each phase being very large, the equivalence between the Kawasaki and Wolff dynamics seems to hold. The agreement is then better than expected.
	}
 	\label{fig:comparison2}
\end{figure}

To check the validity of the effective equilibrium approach, we carried out numerical simulations on small systems, comparing the density profile obtained using the Kawasaki and Wolff methods (see Sec.~\ref{sec:numerical methods} of the SM). The force strength $f$ needed for the Wolff dynamics was chosen separately for each temperature by fitting the current measurements of the Kawasaki dynamics, shown in Fig.~\ref{fig:current}(d).
As is evident from the density profiles given in Figs.~\ref{fig:comparison1}-\ref{fig:comparison2}, the comparison shows good agreements across the temperature range. For larger systems with non-zero magnetization below criticality, we expect only qualitative agreement between the two methods, as the mapping to equilibrium breaks down. 

\section{Effective magnetic field for finite periodic systems}\label{sec:periodicity}

The effective magnetic field discussed in the paper was derived for an infinite system. In two dimensions, when the pump is oriented in the $\hat{x}$ direction ${\bf f}=\hat{x}f$, the field takes the form
\begin{equation}
    h_{\rm eff}(x,y) = \frac{f}{2\pi}\frac{x}{x^2+y^2}\;.
\end{equation}
We remind the reader that this expression is merely the inner product of the gradient of the Green's function of the Laplacian and the force ${\bf f}$ exerted by the pump.

To obtain the magnetic field in a finite square periodic system of dimensions $L\times L$, one needs to derive the Green's function of the Laplacian in this domain.
This can be done in two steps: first, one can obtain the Green's function of a dipole at the origin, pointing in the $\boldsymbol{\hat{x}}$ direction between two \emph{infinite} walls situated at $x=\pm L/2$. This readily gives
\begin{equation}
    h_{\parallel}(x,y;L) = \frac{f \sin\left(\frac{2\pi x}{L}\right)}{\cosh\left(\frac{2\pi y}{L}\right)-\cos\left(\frac{2\pi x}{L}\right)}\;,
\end{equation}
ensuring periodic boundary conditions along $\boldsymbol{\hat{x}}$.
Second, using the expression for the magnetic field between the infinite walls, one can obtain the magnetic field on the square domain. To this end, we construct an infinite array of pumps located at $(x,y)=(0,Ln)$, with $n\in\mathbb{Z}$.
Using Mathematica we obtained the following expression for the density profile:
\begin{align}\label{eq:magnetic field periodic}
    h_{\rm eff}(x,y;L) =& -i\frac{f}{L}\left[\frac{8 \pi x}{L} + \psi_q\left(1+i\frac{x-i y}{L}\right)\right. \nonumber \\
    & -i \psi_q\left(1-i\frac{x+iy}{L}\right)+i \psi_q\left(i\frac{x+iy}{L}\right)\nonumber \\
    &\left.-i \psi_q\left(-i\frac{x-i y}{L}\right)\right]\;
\end{align}
where $\psi_q(z)$ is the $q$-digamma function defined by
\begin{equation}
    \psi_q(z)\equiv-\log(1-q)+\log(q)\sum_{n=0}^\infty \frac{q^{n+z}}{1-q^{n+z}}\;,
\end{equation} 
and $q = e^{2\pi}$. We stress here that the above expression is purely real.

\section{The behavior just above the critical point}

In the analysis carried out in the main text we consider a pump exactly at the critical point where the correlation length $\xi_0$, in the absence of the pump, is infinite. In this Section we discuss the behavior near the critical point and for $T \gtrsim T_c$. We start by considering dimensions $d<4$.

\begin{figure}
 	\centering
	\includegraphics[width=8.6 cm]{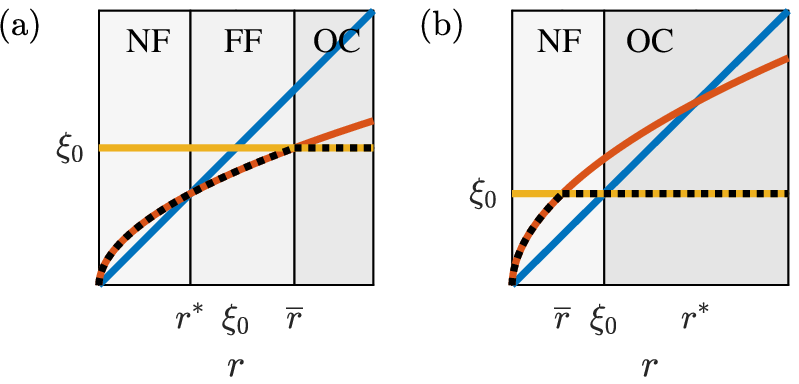}
	\caption{The different length scales $\xi_0$ (orange), $\xi(h_{\rm eff})$ (red), and $r$ (blue) the distance from the pump as function of $r$ in the two scenarios described in the text (a) $\xi_0>r^*$,
	(b) $\xi_0<r^*$. The correlation length that controls the behavior of the system is marked by the black dotted line. The regions marked in shades of gray describe the behavior of the system -- `NF' for critical near-field, `FF' for critical far-field, and `OC' for off-critical high-temperature behavior. The red curve $\xi(h)$ is drawn for $d=2$, having the same qualitative behavior as that of $d=3$.}
	
 	\label{fig:scales}
\end{figure}

Off criticality the thermal correlation length $\xi_0\sim \tau^{-\nu}$ is finite. We expect that on large length-scales the $\tau>0$ behavior of the main text is recovered. More precisely, $\phi^{\rm OC}({\rm r})\sim \chi h_{\rm eff}({\rm r})$ where $\chi$ is the magnetic susceptibility which was given in the main text by $\tau^{-1}$ but close to the critical point is $\chi\sim\tau^{-\gamma}$ where $\gamma=7/4,\,1.24$ in two and three dimensions, respectively. However, one also expects a regime where a behavior similar to the critical behavior of the main text applies. 

Recall that we showed that at $T=T_c$ the correlation length is finite as a result of the magnetic field and it depends on the distance $r$ from the pump. The interplay between the correlation length and the rate of change of $h_{\rm eff}({\bf r})$ dictated whether the far-field or near-field behavior are exhibited. In particular, when the correlation length at a distance $r$ from the pump is larger than $r$ the density is controlled by the near-field behavior.

To understand the behavior of the order parameter on different length scales, we first have to identify the correlation length of the system as a function of $r$. This depends on $h_{\rm eff}({\bf r})$ as well as the deviation of the temperature from the critical value, $\tau$. As a qualitative approximation, it is the minimum of the correlation length induced by either the magnetic field, $\xi(h_{
\rm eff}(\vec{r}))$, or by a nonzero $\tau$, $\xi_0$.

To proceed we plot in Fig.~\ref{fig:scales} $\xi_0,\,\xi(h_{\rm eff})$ and $r$ as functions of $r$ using $\xi(h_{\rm eff}) \sim h_{\rm eff}^{-\frac{(\delta-1)}{\delta(2-\eta)}}$. The three functions intersect at three points. As in the main text we denote by $r^*$ the intersection of $r$ with $\xi(h_{\rm eff})$. Another important intersection point, denoted by $\overline{r}$, occurs when $\xi(h_{\rm eff})=\xi_0$.

Consider first the case depicted in Fig.~\ref{fig:scales}(a) where $r^*<\overline{r}$. This occurs when the pump is strong or $\tau$ is very small,
specifically,
\begin{equation}\label{eq:condition}
    \tau^{\nu(1-\frac{d}{\delta+1})}<f\;.
\end{equation}
In this case when $r<\overline{r}$ the correlation length which controls the system is $\xi(h_{\rm eff})$. Then in this region the system behaves like a critical system with a crossover from the near-field to the far field behavior at $r^*$. For distances $r>\overline{r}$ the correlation length $\xi_0$ controls the fluctuations, resulting in a behavior similar to that of the $\tau>0$ regime of the main text. 
In particular, we have $\phi^{\rm OC} \sim \chi h_{\rm eff}({\bf r})$, as mentioned above. Indeed, it is easy to check the self-consistency of this argument by verifying that $\phi^{\rm OC} \sim  \phi^{\rm nonlinear}$ at $r=\overline{r}$ using the scaling relation $\gamma=\nu(2-\eta)$~\cite{kardar2007statistical}.

Next, consider the case depicted in Fig.~\ref{fig:scales}(b) where $r^*>\overline{r}$, opposite to the condition in Eq.~\eqref{eq:condition}. In this case the correlation length on a scale $r$ is given by  $\xi(h_{\rm eff})$ for $r<\overline{r}$ and by $\xi_0$ for $r>\overline{r}$. Note that this implies that for $r<\xi_0$ the correlation length at a distance $r$ from the pump is larger than $r$ so that the near-field behavior is observed. For $r>\xi_0$ again the correlation length can be ignored and one recovers the $\tau>0$ off-critical behavior of the main-text. The self-consistency of the argument can by checked by verifying that $\phi^{\rm OC} \sim \phi^{\rm linear}$ at $r=\xi_0$. Interestingly, when Eq.~\eqref{eq:condition} is satisfied, there are three different regimes, but only two when it is not.

For dimensions $d>4$, the situation simplifies as there are no near and far-field regimes. Then on scales $r$ larger than $\xi_0$ the density profile matches its off-critical high-temperature form $\phi\sim \vec{f}\cdot\vec{r}/r^{d}$ while for $r<\xi_0$ critical mean-field solution is valid with $\phi\sim \vec{f}\cdot\vec{r}/r^{d-2}$.

\section{Non-zero magnetization below $T_c$}

\begin{figure}
 	\centering
	\includegraphics[width=8.6 cm]{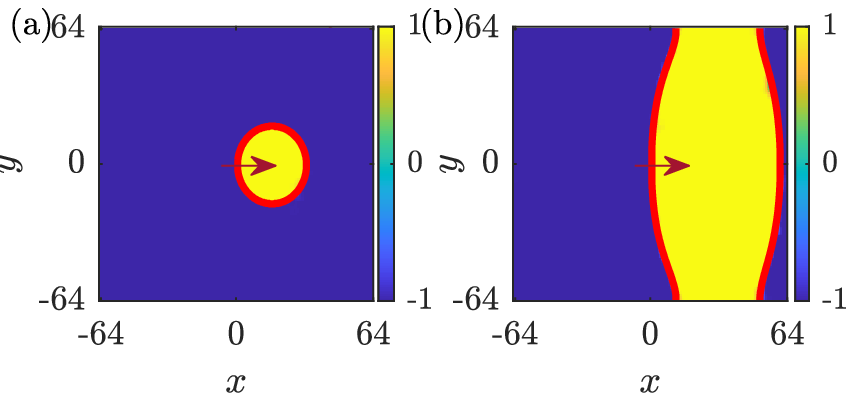}
	\caption{Domain walls created in the effective equilibrium system at a temperature below the critical point $T=0.3T_c$, for two values of the offset magnetic field $h_0$ controlling the overall magnetization $\langle\phi\rangle$. Density profiles, simulated using the Wolff algorithm on a $128\times128$ lattice, are compared with the theoretical domain walls (red lines) predicted by Eq.~\eqref{eq:phase boundary}. Note that due to the breakdown of the mapping to effective equilibrium, these domains do not faithfully describe the original pumped lattice gas (see Fig.~\ref{fig:comparison2}).
	(a) Domain wall created for a magnetization of $\langle\phi\rangle=5.7\cdot10^{-2}$ due to an offset magnetic field of $h_0=2.3\cdot10^{-3} f/a$.
	(b) Domain wall created for a magnetization of $\langle\phi\rangle=0.4$ due to an offset magnetic field of $h_0=2.3\cdot10^{-4} f/a$.}
	
 	\label{fig:Wolff bubbles}
\end{figure}

Here we consider the interesting case when the average magnetization in the system is non zero below the critical temperature. While the mapping breaks down in this regime, a qualitative understanding can be inferred by assuming a constant mobility. We start by considering an infinite system. Rather than fixing the magnetization (which is not well-defined for an infinite system), we study the shape of domains by considering the equilibrium between the two phases. To do so we minimize the free-energy with the effective magnetic field $h_{\rm eff}$ to get
\begin{align}\label{eq:mu boundary}
	\langle\mu\rangle({\bf r}) =& K\nabla^2 \langle\phi\rangle - \left(1+\frac{u}{\tau}\langle\phi\rangle^2 \right) \tau \langle\phi\rangle \nonumber \\
	=& \frac{1}{S_d} \frac{{\bf f}\cdot{\bf r}}{r^d} - h_0 \ .
\end{align}
Here we added the field $h_0$ as a Lagrange multiplier which allow for a non-zero magnetization in the system. On large length scales the gradient term can be neglected. To identify the phase boundary, we note that the magnetization can point in either direction when the local magnetic field, including the $h_0$ is zero. Namely, when
\begin{equation}\label{eq:phase boundary}
    h_0 = \frac{1}{S_d}\frac{{\bf f}\cdot{\bf r}}{r^d} = h_{\rm eff}\ ,
\end{equation}
Then the magnetization along the domain boundaries is $\langle\phi\rangle^2=-\frac{\tau}{u}$. In particular, in two dimensions, after setting  ${\bf f}=\hat{x}f$, Eq.~\eqref{eq:phase boundary} reads
\begin{equation}\label{eq:circles}
	 \left(x-\frac{f}{4\pi h_0}\right)^2 + y^2 = \left(\frac{f}{4\pi h_0}\right)^2\ ,		
\end{equation}
so that a circular domain forms, passing through the pump. As in the case with zero magnetization, the pump localizes the domain wall. For a finite size system, when the radius of the domain reaches $\sim L/4$ this result cannot hold anymore. However, as mentioned in Sec.~\ref{sec:periodicity} of the SM, it is straightforward to extend our results to account for, say, periodic boundary conditions when deriving the expression for $h_{\rm eff}$~\eqref{eq:magnetic field periodic}. The results are compared with the simulated density profiles both for Kawasaki and Wolff simulations in Fig.~\ref{fig:comparison2}.
Note that the deviations from the theory stem from two origins.  First, the discrepancy between the Kawasaki dynamics and the Wolff algorithm are expected because of the breaking down of the mapping to equilibrium. The discrepancy between the theory and the Wolff algorithm is largely due to finite-size effects and the width of the domain walls (the agreement becomes very good for larger systems, see Fig.~\ref{fig:Wolff bubbles}). Nonetheless, the overall qualitative picture is correct.

\bibliography{main}